\documentclass[%
 aip,
 amsmath,amssymb,
reprint,%
]{revtex4-1}
\usepackage{color}
\usepackage{graphicx}
\usepackage{dcolumn}
\usepackage{bm}
\usepackage[utf8]{inputenc}
\usepackage[T1]{fontenc}
\usepackage{mathptmx}
\usepackage{float}

\begin{document}

\preprint{AIP/123-QED}

\title{Suspended gallium arsenide platform for building large scale photonic integrated circuits: Passive devices}

\author{Pisu Jiang}
 \affiliation{Quantum Engineering Technology Labs and Department of Electrical and Electronic Engineering, University of Bristol, Woodland Road, Bristol BS8 1UB, UK}
\author{Krishna C. Balram}
 \email{krishna.coimbatorebalram@bristol.ac.uk}
\affiliation{Quantum Engineering Technology Labs and Department of Electrical and Electronic Engineering, University of Bristol, Woodland Road, Bristol BS8 1UB, UK}
    
\date{\today}

\begin{abstract}
The spectacular success of silicon-based photonic integrated circuits (PICs) in the past decade naturally begs the question of whether similar fabrication procedures can be applied to other material platforms with more desirable optical properties. In this work, we demonstrate the individual passive components (grating couplers, waveguides, multi-mode interferometers and ring resonators) necessary for building large scale integrated circuits in suspended gallium arsenide (GaAs). Implementing PICs in suspended GaAs is a viable route towards achieving optimal system performance in areas with stringent device constraints like energy efficient transceivers for exascale systems, integrated electro-optic comb lasers, integrated quantum photonics, cryogenic photonics and electromechanical guided wave acousto-optics.
\end{abstract}

\maketitle
The scale, complexity and performance of silicon photonic integrated circuits (PICs) has revolutionized optical communications in the past decade \cite{thomson2016roadmap}. Perhaps the most surprising aspect of this revolution is the fact that silicon does not possess many desirable optical properties (apart from a high refractive index) and the silicon photonics revolution was primarily driven by the availability of a foundry fabrication infrastructure, courtesy of the microelectronics industry, that could be applied to optics \cite{hochberg2010towards,atabaki2018integrating,giewont2019300}. Over the past two decades, a wide variety of component designs have been optimized and their fabrication process perfected for silicon \cite{chrostowski2015silicon} and it is hard to foresee a similar investment of resources in any other material platform. On the other hand, there are a number of application areas in which silicon's lack of desirable optical properties proves a severe limitation to achieving system performance. These limitations include the absence of a direct bandgap, lack of a $\chi^{(2)}$ nonlinearity to build fast electro-optic devices and zero piezoelectric response which makes it challenging to design acousto-optic devices. As a representative example, one of the key challenges facing transceivers for exascale systems \cite{shen2019silicon} is avoiding the $\sim$ 3 dB penalty for coupling light from the III-V laser die to the silicon PIC. Electro-optic frequency comb \cite{zhang2019broadband} based coherent communication systems \cite{pfeifle2014coherent} will also benefit greatly from monolithic integration of lasers and modulators. On the quantum photonics side, one of the outstanding problems facing linear optic implementations of quantum computing is implementing feed-forward routines on a chip \cite{mendoza2016active}, which requires (monolithically) interfacing fast, low loss modulators with efficient single photon detectors. Despite the outstanding performance improvements of carrier based depletion modulators, electro-optic modulators present the only near-term solution  that can satisfy both the bandwidth ($\sim$ 40 GHz) and loss requirements (< 3 dB) necessary for scalability \cite{wang2018integrated}. Other application areas where alternative material platforms are worth exploring are: cryogenic photonic circuits for interfacing superconducting digital circuits with the outside world \cite{sobolewski2001ultrafast} and integrated acousto-optics, which requires a piezoelectric material for exciting acoustic waves \cite{de2005modulation,hatanaka2014phonon,balram2016coherent}.    

Gallium arsenide (GaAs) presents a viable alternative to silicon for these applications as it possesses all the desirable optical properties that silicon lacks: a direct bandgap, a $\chi^{(2)}$ nonlinearity, and a (weak) piezoelectric coefficient \cite{dietrich2016gaas}. More importantly, GaAs has a refractive index that is almost identical to silicon ($\lambda$ = 1.55 ${\mu}$m), making it easy to port a variety of optimised photonic designs and fabrication process flows between the platforms. In contrast to other electro-optic platforms like lithium niobate, it has a higher refractive index allowing compact component design, which is key to monolithic systems integration. GaAs also provides a natural route towards incorporating active gain media like quantum dots and wells, which are promising for applications in both classical and quantum \cite{lodahl2015interfacing} photonics. Traditionally integrated photonics in GaAs has suffered from the low index contrast achievable between GaAs and the AlGaAs buffers which serve as waveguide cladding layers. The low index contrast leads to large mode sizes and bend radii which make photonic integration challenging \cite{dietrich2016gaas}. In addition, the reduced optical power density (due to larger mode area) makes it difficult to adequately exploit the nonlinear coefficients for frequency conversion and EO modulator applications \cite{wang2014gallium}. In recent years, there has been tremendous progress in the development of thin films of GaAs on low index media (particularly silicon oxide and nitride) by wafer bonding \cite{zhang2019iii} and a wide variety of devices showing impressive nonlinear performance have been demonstrated \cite{pu2016efficient,chang2019strong,Chiles:19} . On the other hand, wafer bonding is well-known to be a notoriously fickle process and it is challenging to get high device yields (a prerequisite for building PICs) in an academic cleanroom environment. 
 
\begin{figure}[!b]
    \centering
    \includegraphics{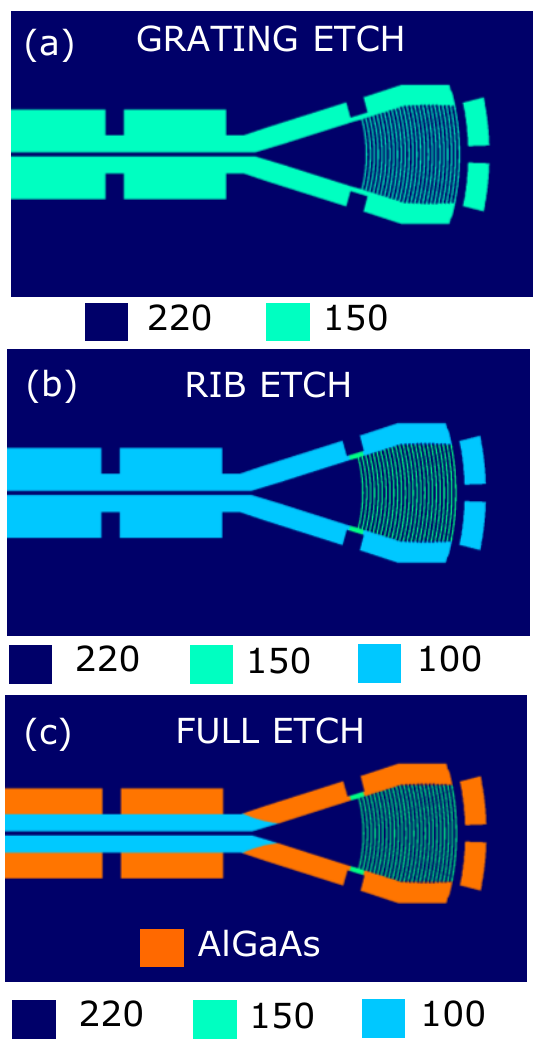}
    \caption{Illustrative schematic of the process flow for building a suspended waveguide platform in GaAs. The process follows the standard three step process used to fabricate passive silicon photonic devices. The GaAs thickness (in nm) in each region is indicated.  (a) A 70 nm GRATING etch is used to define the grating coupler in GaAs. (b) The GaAs is etched a further 50 nm to define the rib waveguides (RIB etch). (c) Finally, a 100 nm FULL etch is carried out to reach the AlGaAs layer. The exposed AlGaAs layer is selectively etched away in weak HF acid and the sample is covered with $\sim$ 2 ${\mu}m$ SiO$_{2}$.}
    \label{fig:1_processflow}
\end{figure}

In this work, we show that a multi-step fabrication process, derived from a standard passive silicon photonics platform  \cite{littlejohns2019rapid}, can be applied to build large scale photonic integrated circuits in suspended GaAs. The suspension of the GaAs layer, achieved by selective etching of the underlying AlGaAs film, is necessary to achieve high refractive index contrast \cite{stievater2015suspended}. By moving to a tethered rib waveguide geometry and judicious choice of etch release holes, all the components of a standard (passive) PIC platform, in particular grating couplers, waveguides, ring-resonators and waveguide splitters (multi-mode interference couplers), can be adapted to the suspended GaAs platform without compromising performance efficiency . The passive devices reported in this work serve as a key building block for the development of active devices, in particular, efficient integrated electro-optic and acousto-optic modulators, which are currently under development. 

\begin{figure}[!b]
    \centering
    \includegraphics[width = \linewidth]{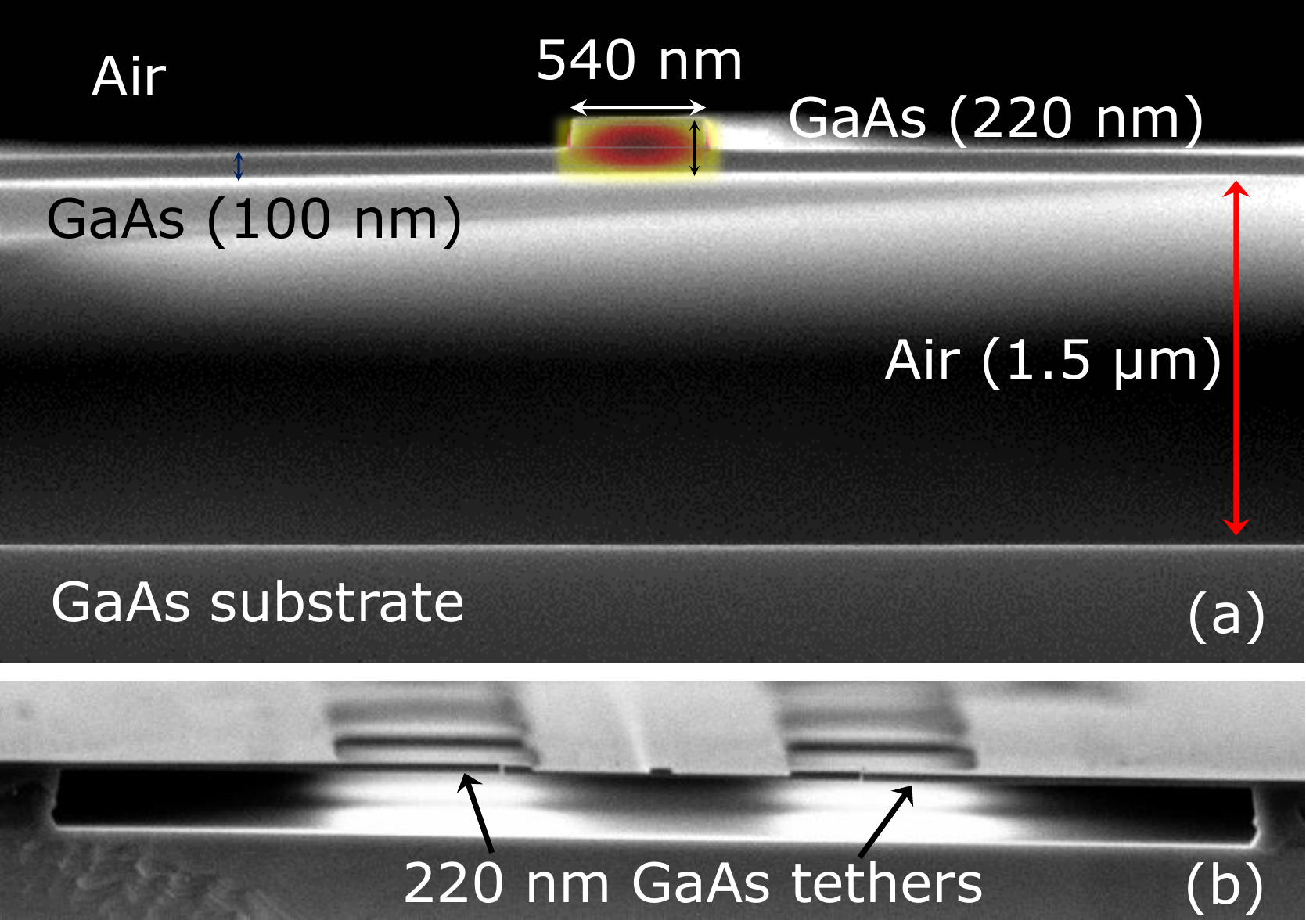}
    \caption{(a) Cross-section SEM image of a suspended GaAs rib waveguide structure. The electric field ($|E_{norm}|$) of the propagating transverse electric (TE) mode, calculated using a numerical FEM simulation using COMSOL, is overlaid. (b) Zoomed-out cross-section showing the GaAs rib waveguide held by suspension tethers. The extent of the HF undercut can be clearly seen.}
    \label{fig:2_ribwvg}
\end{figure}

Figure \ref{fig:1_processflow} shows a schematic illustration of the main process steps for a grating coupler fabricated using this process. The process starts by patterning the grating coupler teeth and defining the outline of the waveguides and the tethers (GRATING etch) in GaAs (Fig. \ref{fig:1_processflow} (a)). The patterning is carried out using electron beam lithography with hydrogen silsesquioxane (HSQ) resist and the GaAs layer is etched using a standard Ar/Cl$_{2}$ chemistry with etch thickness monitored using an ellipsometer to ensure precise etch depths are achieved. This is followed by a RIB etch step (Fig. \ref{fig:1_processflow} (b)), where the GaAs layer is etched a further 50 nm to define the rib waveguides. The grating coupler region is protected with HSQ resist during this step. The two layers are registered with respect to each other using a set of alignment marks defined during the GRATING etch step. A FULL etch step is next carried out by etching the remaining 100 nm of the GaAs layer (+ 20 nm overetch) to access the AlGaAs buffer as shown in Fig.\ref{fig:1_processflow}(c). To suspend the GaAs layer, the AlGaAs buffer is selectively etched in dilute hydrofluoric acid (HF) solution. To remove any remnants of etch residue, the sample is cleaned in a dilute potassium hydroxide solution and flash dried using isopropanol \cite{midolo2015soft}.

\begin{figure}[!h]
    \centering
    \includegraphics[width = \linewidth]{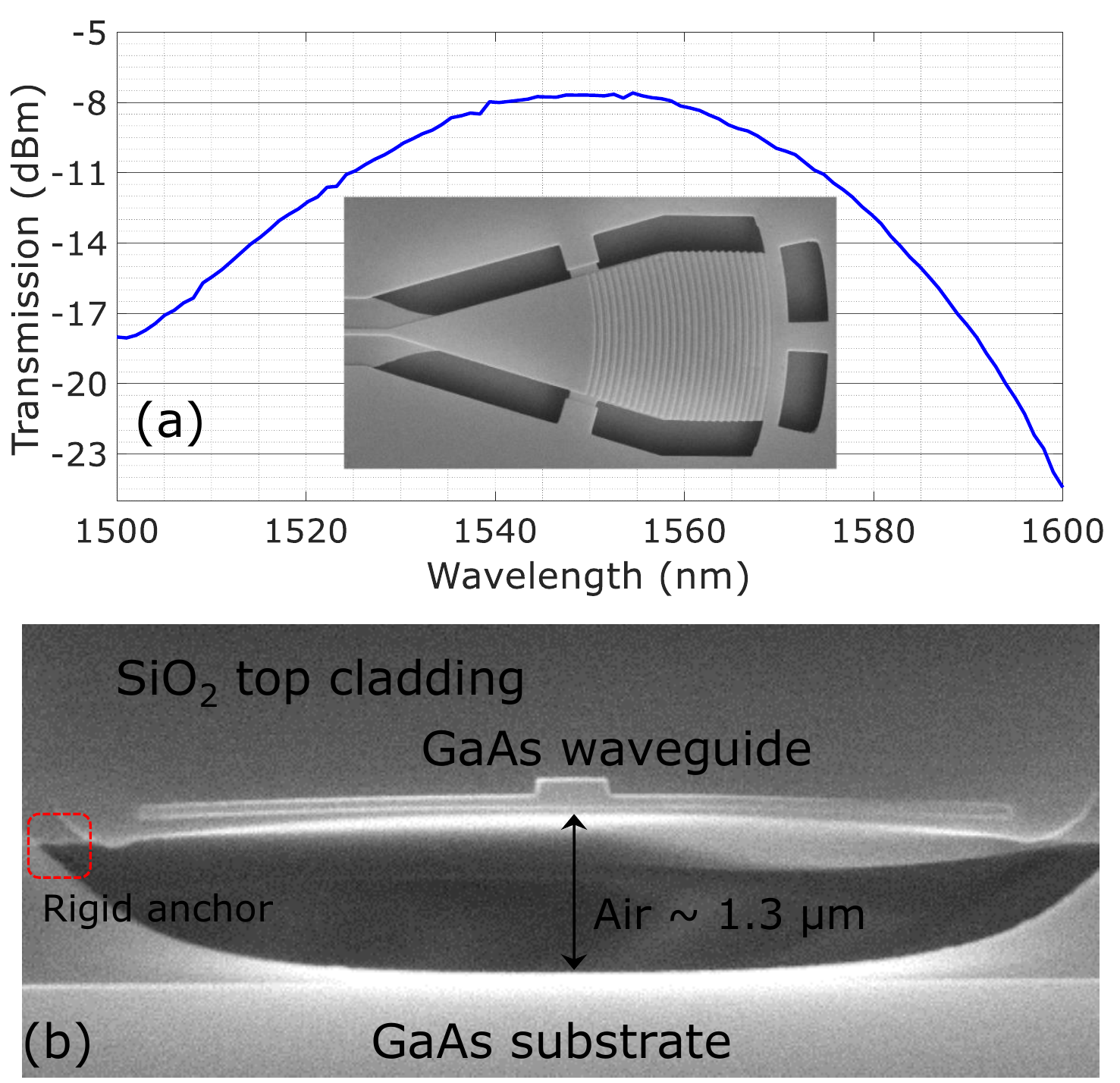}
    \caption{(a) Measured fiber to fiber transmission spectrum of a representative grating coupler with parameters, period ($\Lambda$) = 660 nm, duty cycle ($\eta$) = 0.5 and etch depth ($d$) 70 nm. The fiber array was angled at 12 deg. and the coupling was optimised for maximum transmission. An SEM image of a fabricated grating coupler before oxide encapsulation is shown in the inset. (b) Cross-section image of a GaAs rib waveguide after silicon oxide encapsulation, showing the bottom air gap is reduced to $\sim$ 1.3 ${\mu}m$. The rigid anchoring of the structure at the ends, due to the oxide encapsulation, is indicated by the red box.} 
    \label{fig:3_GC}
\end{figure}

A cross-section of a suspended GaAs rib waveguide fabricated in this platform is shown in Fig.\ref{fig:2_ribwvg}(a). A zoomed-out SEM image of the rib waveguide suspended by tethers is shown in Fig.\ref{fig:2_ribwvg} (b). The normalized total electric field of the fundamental transverse electric (TE) mode, calculated using a numerical mode solver (COMSOL Multiphysics), is overlaid for reference. The high refractive index of the GaAs layer ensures that the mode is mainly confined to the rib region, with very little leakage into the surrounding GaAs or the AlGaAs buffer. By designing rib waveguides with waveguide width $\sim$ 540 nm and total rib width $\sim$ 6 ${\mu}$m, we can ensure that the optical field has negligible overlap ($\eta_{ov}$) with the remaining AlGaAs buffer layer and is tightly confined within the GaAs waveguide. The choice of the rib width was mostly determined by the difficulty of suspending wider GaAs devices on account of the lack of intrinsic stress in the GaAs layer. As mentioned before, the high refractive index allows us to work with smaller rib widths without sacrificing performance (waveguide loss). More importantly, it allows us to design compact, low loss suspended waveguide bends with bend radii $\sim$ 25 ${\mu}$m in this work, and the prospect of achieving bend radii $\approx$ 5 ${\mu}m$. After HF release, the suspended GaAs film is capped with $\sim$ 2 ${\mu}m$ of silicon oxide deposited using plasma enhanced chemical vapor deposition (PECVD). The oxide film is necessary for separating the metal electrodes (required for the electro-optic devices) from the GaAs layer. In addition, they provide mechanical rigidity to the suspended films by pinning them at the corners of the etch holes (shown by the red box in Fig.\ref{fig:3_GC}(b)). To test this rigidity, we performed a stress test by placing the chip (in isopropanol) in an ultrasonic bath at full power for $>$ 10 minutes. After removal from the bath, the chip showed no signs of weakening or structural damage, which ensures that it can survive later processing, critical for active devices. We are able to land fiber arrays on the GaAs chip without noticable structural damage to the devices during optical characterisation. The mechanical rigidity is key for enabling the deposition of thick metallic electrodes for active devices and provides great benefit from a packaging perspective, which is especially critical for cryogenic operation.

The standard set of passive design components available as part of a standard silicon photonics process development kit (PDK) are low-loss waveguides, grating couplers (GC), resonators and on-chip waveguide splitters and combiners. Amongst these, the grating coupler is probably the most important component, as it serves as the interface between the PIC and the outside world. A compact low-loss grating coupler is indispensable for large scale PICs as it allows in-situ device characterisation, without the need for chip-cleaving. Traditionally, GaAs based devices have relied on edge coupling as it is challenging to design efficient grating couplers when the index contrast between the core and cladding is low. While compact free space grating couplers have been optimized by the quantum dot community \cite{zhou2018high}, their coupling efficiency is low and they are not suitable for building large scale PICs. On the other hand, efficient out-of-plane grating couplers have long served as the de-facto standard in the silicon photonics community. Fig.\ref{fig:3_GC}(a) plots the measured fiber-fiber transmission spectrum of a grating coupler test structure that consists of two grating couplers linked by a suspended ridge waveguide. The device was probed using a fiber array angled at 12 deg. The separation between the couplers is 127 ${\mu}$m and the waveguide bends have a radii of 25 ${\mu}$m. An SEM image of our suspended GaAs focusing grating coupler is shown in the inset of Fig.\ref{fig:3_GC}(a). The grating coupler is supported using tethers, defined during the GRATING etch step. The high refractive index of GaAs allows us to keep the light confined to the central region ($t_{GaAs}$ = 220 nm) and efficiently focus it into the rib waveguide. The critical fabrication step in suspending a grating coupler is ensuring that all of the GC is released during the wet etch step. This is ensured by providing release holes both to the side and rear of the GC, as can be seen in the inset. The grating coupler design is modified from the standard silicon designs to account for the difference in surrounding refractive indices and thicknesses. In the suspended GaAs devices, the underside cladding is air and the topside cladding is silicon dioxide. This helps reduce the insertion loss, as the guided wave is more effectively scattered towards the higher index side (the top oxide) towards the fiber, rather than towards the substrate. The main parameters affecting the coupling efficiency are the grating period ($\Lambda$), etch depth ($d$), duty cycle ($\eta$), and thickness of the AlGaAs cladding layer. For $\lambda$ = 1550 nm, the design parameters used are  $\Lambda$ = 660 nm, $d$ = 70 nm, ${\eta}$ = 0.5. The peak fiber to fiber coupling efficiency is $\sim$ -8 dB at $\sim$ 1550 nm, which amounts to $\sim$ 4 dB insertion loss per coupler. One of the key parameters that affects the optimal coupling efficiency is the gap between the GaAs device layer and the substrate, which is determined initially by the AlGaAs thickness. The starting AlGaAs thickness of 1.5 um lies at the bottom of the coupling efficiency curve and any reduction in the spacing between the GaAs membrane and the substrate will improve that number. As can be seen from a zoomed-in cross-section SEM image of the waveguide in Fig.\ref{fig:3_GC}(b), the deposition of the top cladding oxide loads the membrane making it sag. This lowers the gap between the membrane and the substrate and ends up increasing the coupling efficiency. From our cross-section SEM images, we currently estimate the bottom gap to be $\sim$ 1.3 ${\mu}$m.

\begin{figure}[!h]
    \centering
    \includegraphics[width = \linewidth]{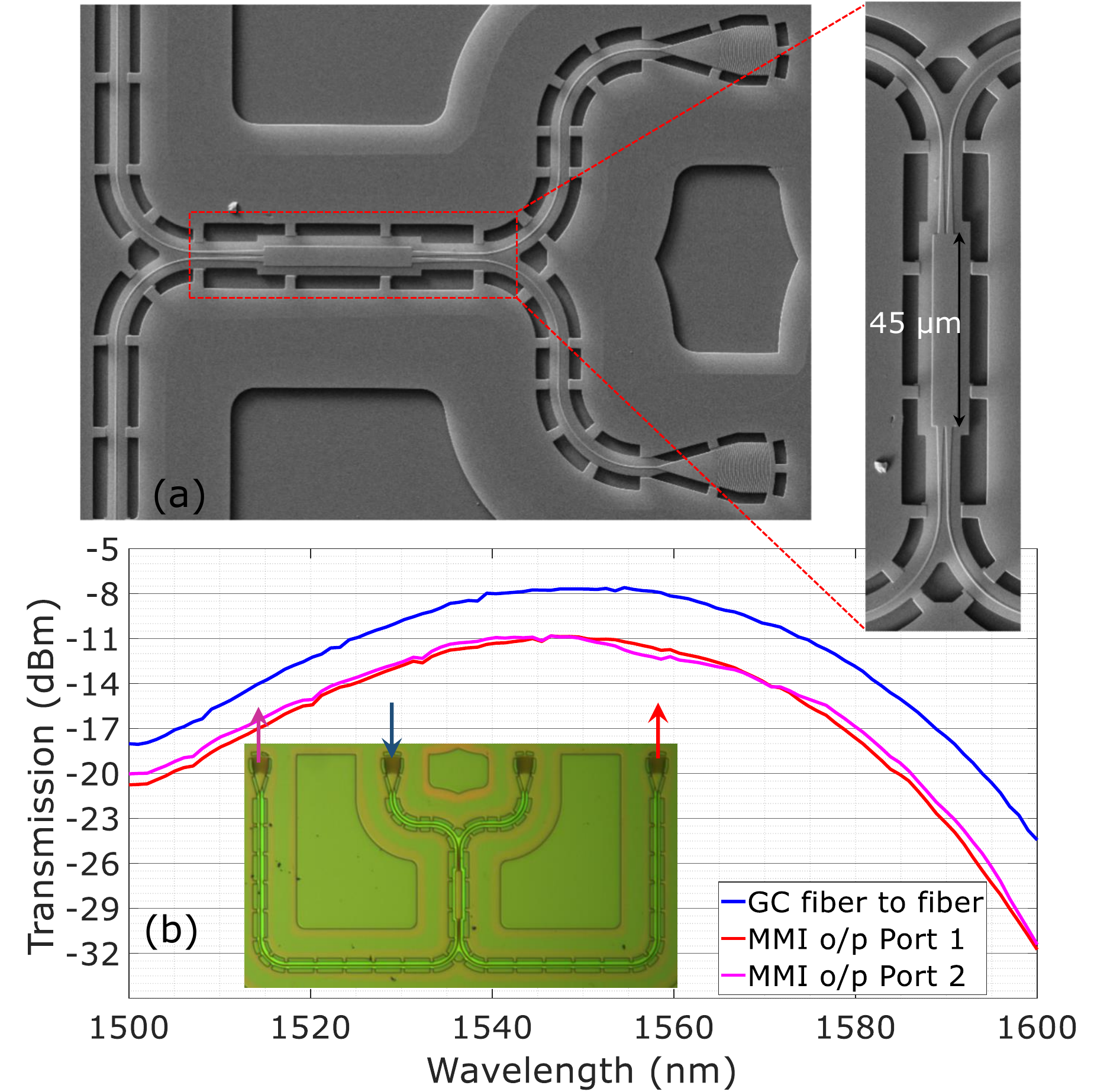}
    \caption{(a) SEM image of a fabricated suspended 2x2 multimode interference (MMI) coupler. A zoomed-in image of the MMI is shown on the right. (b)  Measured transmission spectrum through the two output ports (labelled 1 and 2 and indicated by the red and magenta curves). The reference fiber to fiber transmission spectrum (from Fig. \ref{fig:3_GC}(a)), without the MMI, is shown in blue. A microscope image of the complete MMI with the input and output ports labelled is shown in the inset.}
    \label{fig:4_MMI}
\end{figure}

To build scalable PICs, it is critical to have the ability to split and recombine light. Linear networks of waveguide splitters form a key building block for optical implementations of quantum information processing \cite{carolan2015universal}, deep neural networks \cite{shen2017deep} and optical phased arrays \cite{sun2013large}. A 2x2 multimode interference coupler (MMI) is the standard building block that underpins these linear networks. Fig.\ref{fig:4_MMI}(a) shows an SEM image of 2x2 MMI fabricated using the suspended GaAs platform and based on a standard silicon foundry design \cite{littlejohns2019rapid}. A zoomed-in image of the same device is also shown. The measured transmission spectrum of the best-performing MMI is shown in Fig.\ref{fig:4_MMI}(b). The plot shows the transmitted power from the two output ports (labelled by red and magenta arrows in the inset microscope image). The grating coupler transmission spectrum from Fig.\ref{fig:3_GC}(a) is overlaid for reference. From Fig.\ref{fig:4_MMI}(b), it is clear that the excess insertion loss introduced by the MMI (over the 3 dB due to power splitting) is negligible and these devices can be used to effectively split and re-combine light on a GaAs chip, exactly analogous to silicon. Like with the grating coupler, the high index contrast of GaAs enables us to port these designs from silicon to the GaAs platform with minimal re-design. The MMI devices also give a sense of the the scale and complexity of the devices that can be engineered. As silicon photonics and before it silicon microelectronics have shown, once a set of robust building blocks have been demonstrated, circuits of arbitrary complexity can be synthesized by connecting these building blocks together in the desired order. 

\begin{figure}[!h]
    \centering
    \includegraphics[width = \linewidth]{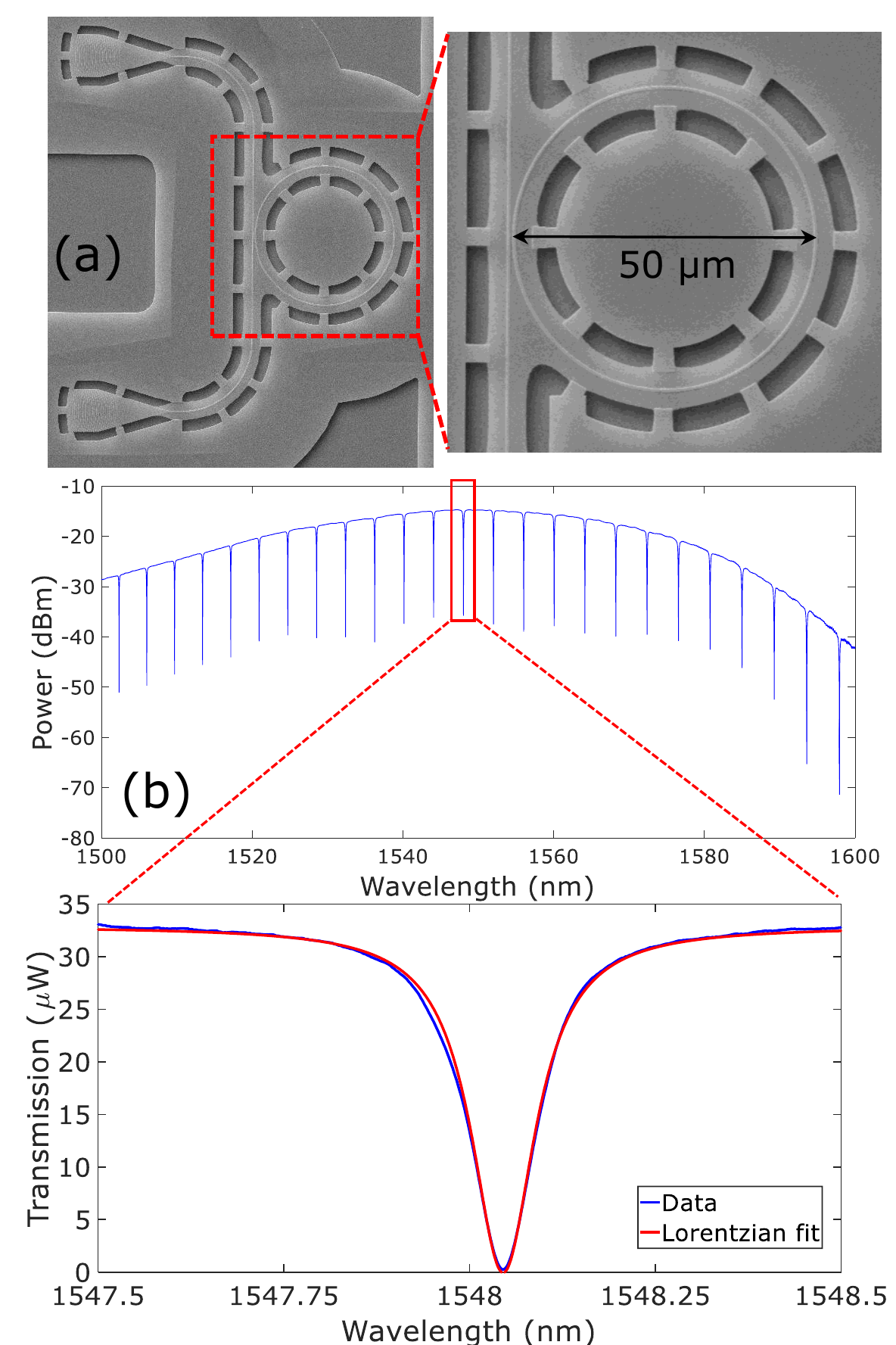}
    \caption{(a) SEM image of a suspended GaAs microring resonator with radius 25 ${\mu}m$ and waveguide resonator gap 325 nm. A zoomed-in image of the ring is shown on the right for reference. (b) Measured transmission spectrum of the resonator showing a series of TE resonant modes ($P_{in}$ = -6 dBm). The TE polarisation is determined by the grating coupler. (c) Zoomed-in scan of one of the resonances, fitted with a Lorentzian lineshape, giving a measured $Q_{opt}\sim$ 15000.}
    \label{fig:5_microring}
\end{figure}

The final component of the passive devices toolkit for suspended GaAs photonics are microring resonators. High quality factor dispersion engineered microring resonators have served as the foundation of a variety of experimental advances in photonic sensing, frequency combs and on-chip generation of single photon pairs by spontaneous four wave mixing \cite{osgood2009engineering}. Fig.\ref{fig:5_microring}(a) shows an SEM image of a suspended microring resonator fabricated using this platform. The ring radius is 25 ${\mu}$m and the waveguide resonator gap was designed to be 325 nm. Fig.\ref{fig:5_microring}(b) shows the measured transmission spectrum of the resonator showing a series of TE resonances. The TE mode selectivity is determined by the polarisation selectivity of the grating coupler.

Fig.\ref{fig:5_microring}(c) shows a narrow band transmission scan around one of the resonances. Overlaid is a fit to the spectrum using a Lorentzian lineshape. The extracted quality factor ($Q_{opt}$) of the device is $\sim$ 15000. While the $Q_{opt}$ is lower than expected, there are several design and fabrication tweaks that can be made to increase it. On the design side, the waveguide width can be increased to provide greater mode confinement. On the fabrication front, we can significantly improve our post-release device cleaning procedures to remove any residual contaminants and incorporate surface passivation techniques that have shown to improve $Q_{opt}$ \cite{guha2017surface}. Based on the measured resonator $Q_{opt}$ and free spectral range (from Fig.\ref{fig:5_microring}(b)), we estimate our current waveguide propagation loss to be $\sim$ 7 dB/cm. We believe the resonator is currently operating in the overcoupled regime, and hence the propagation loss estimate is an upper bound.    

In summary, we have demonstrated that the complexity of the standard (passive) silicon photonics process can be readily transferred to more interesting optical materials, in particular GaAs. The similarity of the refractive indices of the two materials ensures that high device performance can be readily ensured without requiring extensive component re-design. Bringing the scale and complexity of silicon photonics to more interesting optical platforms will be revolutionary for device applications in wide-ranging areas from quantum photonics to cryogenic photonic circuits. Moving forward, we will extend this platform to demonstrate low-loss long spiral ($L_{wvg}{\sim}$ cm) waveguides, and active electro-optic and acousto-optic devices.

\textbf{Acknowledgements}: K.C.B. would like to thank Kartik Srinivasan for lending him the GaAs substrate used for carrying out this work. Sample fabrication was also carried out on GaAs substrates provided by the UK Engineering and Physical Sciences Research Council (EPSRC) National Epitaxy Facility in Sheffield, through a pump-prime grant. We would like to thank Ed Clarke, Imad Faruque, Edmund Harbord, George Kanellos, Laurent Kling, Josh Silverstone, Joe Smith, Kartik Srinivasan, and Siyuan Yu for valuable discussions and suggestions. K.C.B. would like to acknowledge funding support from the European Research Council (ERC-StG, SBS3-5 / 758843). Nanofabrication work was carried out on equipment funded by the EPSRC (EP/N015126/1).

\section{References}

\bibliography{GaAs_refs}

\end{document}